\documentclass[12pt]{article}

\usepackage{geometry,graphicx}

\usepackage{pslatex}
\usepackage{hyperref} 

\usepackage{graphicx}

\newcommand{\delz}{$<\!\delta z\!>$}

\begin{document}
\begin{center}                  
{\large\bf Isolated hard photons with jets measured in Deep Inelastic
Scattering using the ZEUS detector at HERA}
\vspace*{3.0mm}\\ 
Peter J Bussey.\\[3mm]
University of Glasgow, Glasgow G12 8QQ, U.K.\\
E-mail: peter.bussey@glasgow.ac.uk \vspace{2mm}\\
{\em for the ZEUS Collaboration}

\end{center}

\abstract{Isolated hard photons have been 
measured with jets in Deep Inelastic Scattering using the ZEUS
detector at HERA.  Preliminary results for the cross sections are presented.}

~\\[5mm]
Presented at the 2011 Europhysics Conference on High Energy Physics, EPS-HEP 2011,\\
		July 21-27, 2011\\
		Grenoble, Rh\^one-Alpes, France.

\section{Introduction}
Photons with high transverse momentum, $p_T$, may be produced in Deep
Inelastic Scattering of electrons (or positrons) by protons in various
ways.  They may be (i) produced in a hard partonic scattering process,
(ii) radiated from the incoming or the outgoing lepton, (iii) radiated
from a quark that has been produced at high $p_T$, or (iv) a decay
product of a high energy hadron.  Processes (i) and (ii) generate
outgoing photons that tend to be isolated from the other particles in
the final state, and the photons from processes of type (i) are
conventionally known as ``prompt'' photons.  Some typical Feynman
diagrams for processes (i) and (ii) are shown in figure 1.

\begin{figure}[b]
~\\[-12mm]
\begin{center}
\includegraphics[width=.20\textwidth]{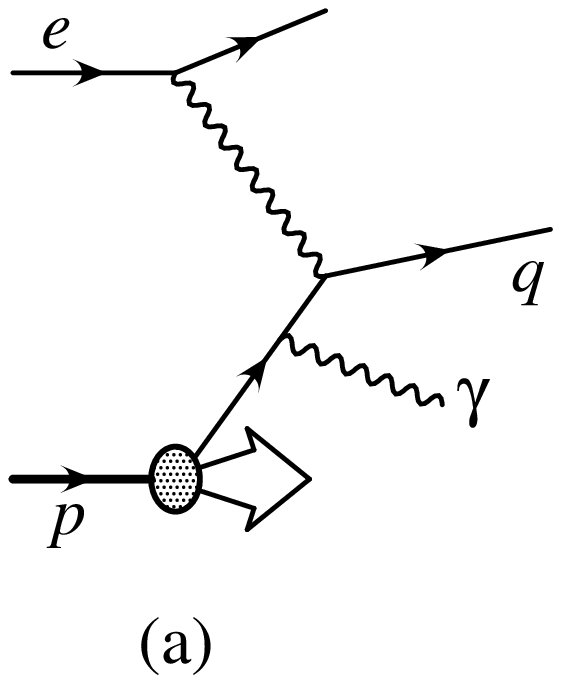}
\hspace{0.04\textwidth}
\includegraphics[width=.20\textwidth]{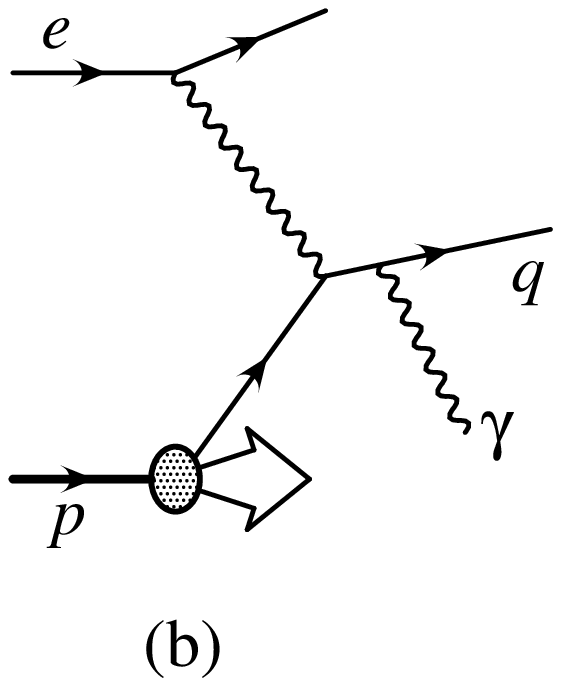}
\hspace{0.04\textwidth}
\includegraphics[width=.20\textwidth]{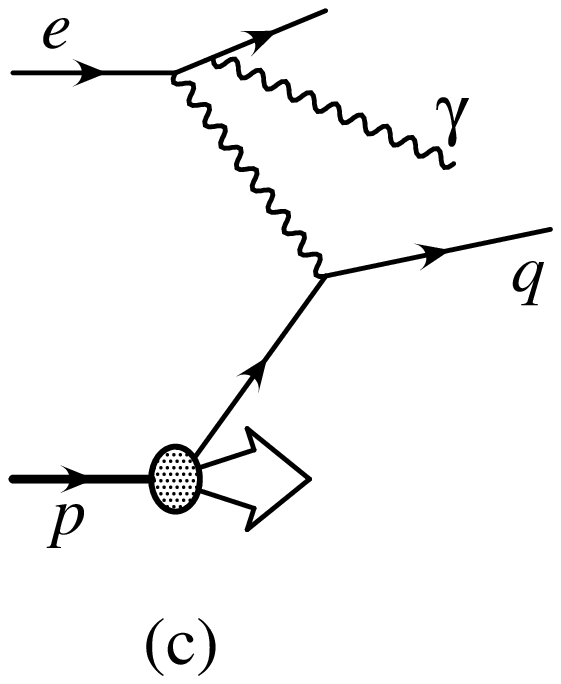}
\hspace{0.04\textwidth}
\includegraphics[width=.20\textwidth]{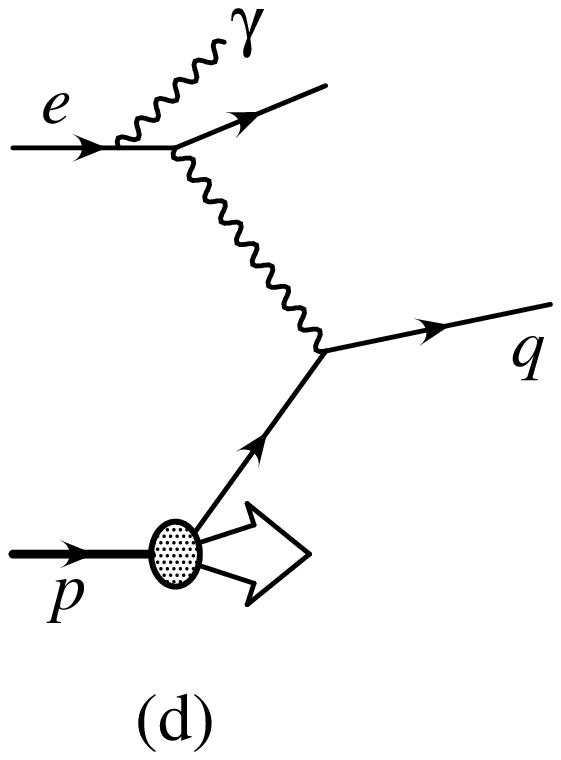}
\\[4mm]
\caption{\small Diagrams for hard photon processes in DIS. Processes
(a), (b) are for ``prompt'' photons, which are radiated from a quark,
referred to here as QQ processes.  In processes (c), (d) the photon is
radiated from a lepton, referred to here as LL processes.}
\end{center}
\end{figure}

These processes are of interest, since they give a
distinctive perspective on QCD physics. In particular, the prompt
photons are produced and detected directly from the basic parton
scatter and are not formed through a jet fragmentation process.
Particular theoretical models can be tested.  In a previous
publication, the ZEUS collaboration presented inclusive measurements
of isolated hard photons in DIS processes~\cite{zeus1}.  Here,
preliminary results are presented for measurements in which a jet is
observed in addition to a hard photon~\cite{zprel}.  This enhances the
prompt component in the data sample.

\section{Apparatus and measurement}
The ZEUS detector operated at the HERA collider, in which 
electrons and positrons at 27.5 GeV collided with protons at 920 GeV.
The principal components of the ZEUS detector used in this analysis were a central
drift-chamber tracker within a solenoidal magnetic field, surrounded
by a uranium-scintillator calorimeter. The calorimeter was divided
into three regions, forward, barrel and rear, and each region
consisted of a finely segmented electromagnetic section outside which
was a hadronic section with larger cells.  ``Forward'' refers to the
proton beam direction.

\begin{figure}[!b]
~\\[-8mm]
\begin{center}
\includegraphics[width=.5\textwidth]{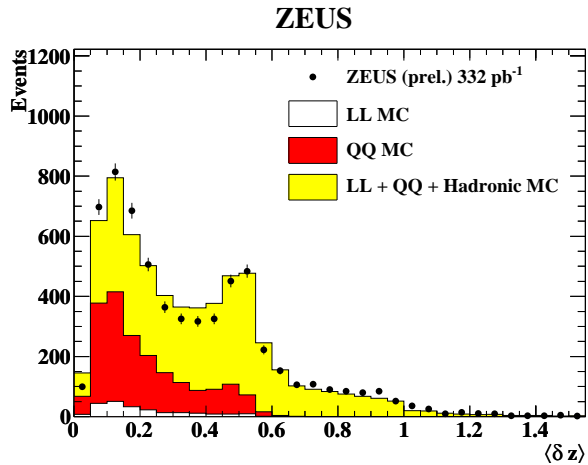}
\\[0mm]
\caption{\small Distribution of the mean longitudinal width measure \delz\  of
the electromagnetic calorimeter cells forming the photon candidate,
fitted to a combination of background, LL photons and QQ photons.
}
\end{center}
\end{figure}

The present analysis uses 332 pb$^{-1}$ of data taken during
2004-2007. A photon candidate in an event corresponds to a
closely-spaced cluster of barrel calorimeter cells that have fired,
giving a total transverse energy $E_T^\gamma$ of at least 4 GeV.  The
pseudorapidity $\eta^\gamma$ of the photon candidate must lie in the
range -0.7 to 0.9, within the barrel calorimeter acceptance.  Jets are
reconstructed from ZEUS energy flow objects~\cite{efo}, which combine
tracking and calorimeter information, and the $k_T$ clustering
algorithm is used~\cite{ktclus}.  The jet must have a transverse
energy $E_{T \mathrm{jet}}$ of at least 2.5 GeV and a pseudorapidity
$\eta_\mathrm{jet}$ in the range -1.5 to 1.8.  The photon candidate
must have at least 90\% of its energy in the electromagnetic
calorimeter cells, and must be isolated in the sense that in the
reconstructed jet-like object containing mainly the photon candidate,
the latter must take at least 90\% of the transverse energy. To reduce
photoproduction background, the scattered beam electron (positron)
must have an energy of at least 10 GeV, must be scattered at an angle
of at least 140$^\circ$ from the proton direction, and must correspond
to a transverse momentum squared, $Q^2$ of between 10 and 350 GeV$^2$.

A substantial background arises from high energy neutral mesons, in
particular $\pi^0$ mesons.  To extract the photon signal, the quantity
\delz\ is used, defined as the $E_T$-weighted mean of the distance of
the $Z$ position of the electromagnetic cells in the cluster from the
mean $Z$ of the cluster. This is illustrated for the entire sample in
figure 2, where the distribution is fitted to the sum of the LL
contribution and a freely scaled background contribution, both
evaluated from the Ariadne 4.12 Monte Carlo~\cite{ariadne}, and a
freely scaled QQ contribution, evaluated using Pythia
6.416~\cite{pythia}.  The photon contributions have a peak at low
\delz\ , plotted in units of electromagnetic calorimeter cell widths,
indicating that most of the energy is found in one cell.  The
background is broader, and peaks around a value of approximately 0.5,
where the cluster energy is mostly divided between two contiguous
cells.  Fits of this kind are performed to the data in each bin of each
quantity whose cross section is to be evaluated, to extract the photon
signal.

\section{Results}
The resulting cross sections as functions of $Q^2$, Bjorken $x$,
$E_T^\gamma$, $\eta^\gamma$, $E_{T \mathrm{jet}}$ and $\eta_\mathrm{jet}$ are shown in
figure 3.  Also shown is the sum of the LL contribution, the QQ
contribution from Pythia, scaled by a factor 1.6, and the Ariadne
generated background, scaled by the same factor. The $Q^2$
distributions of the fitted contributions have been scaled to fit
the data. Systematic uncertainties are dominated by the photon and jet
energy scales, and the modelling of the background. The overall
agreement between the data and this model is very good.\\[5mm]

\begin{figure}[!b]
\vspace{100mm}
\begin{center}
\includegraphics[width=.45\textwidth]{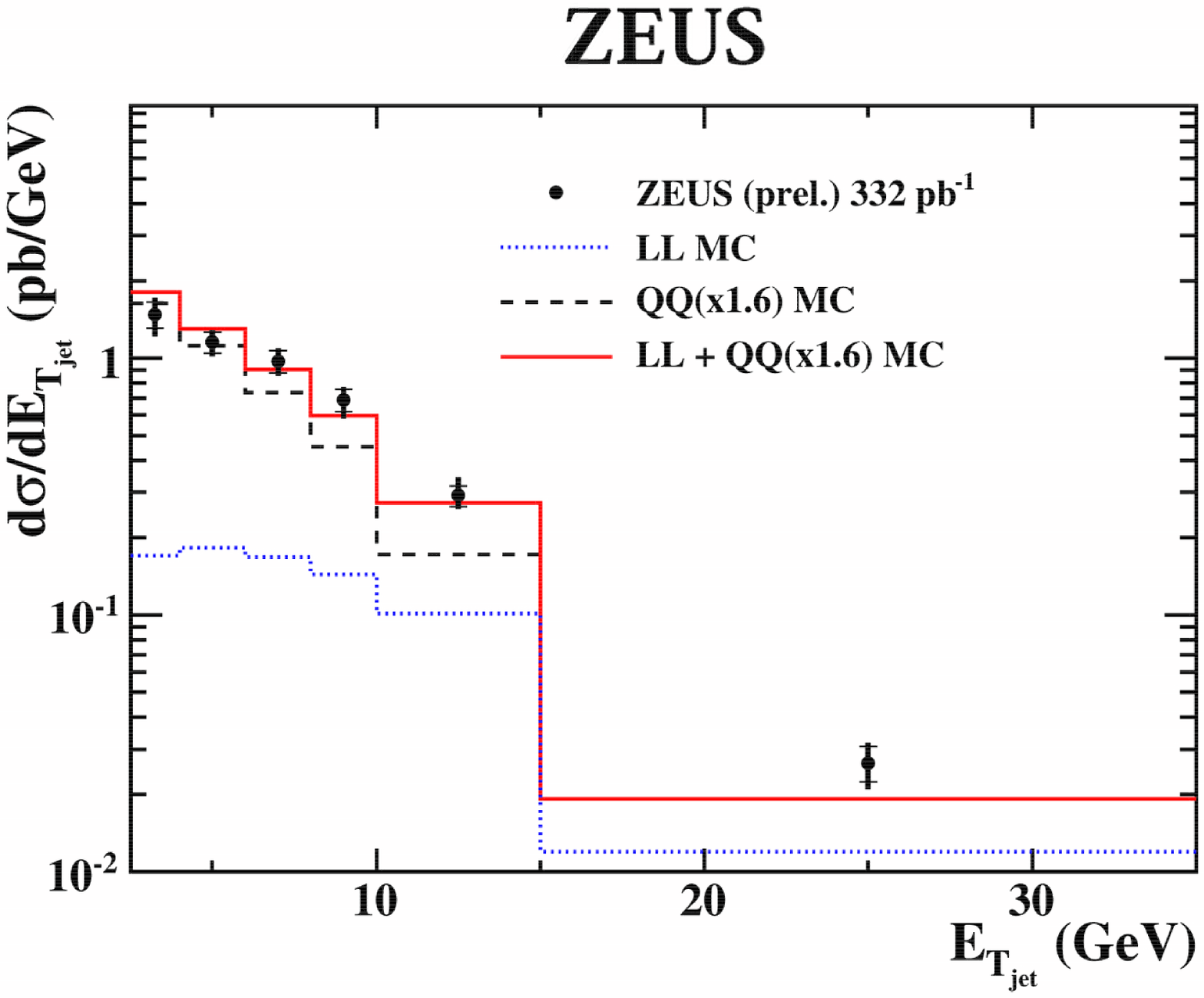}
\includegraphics[width=.45\textwidth]{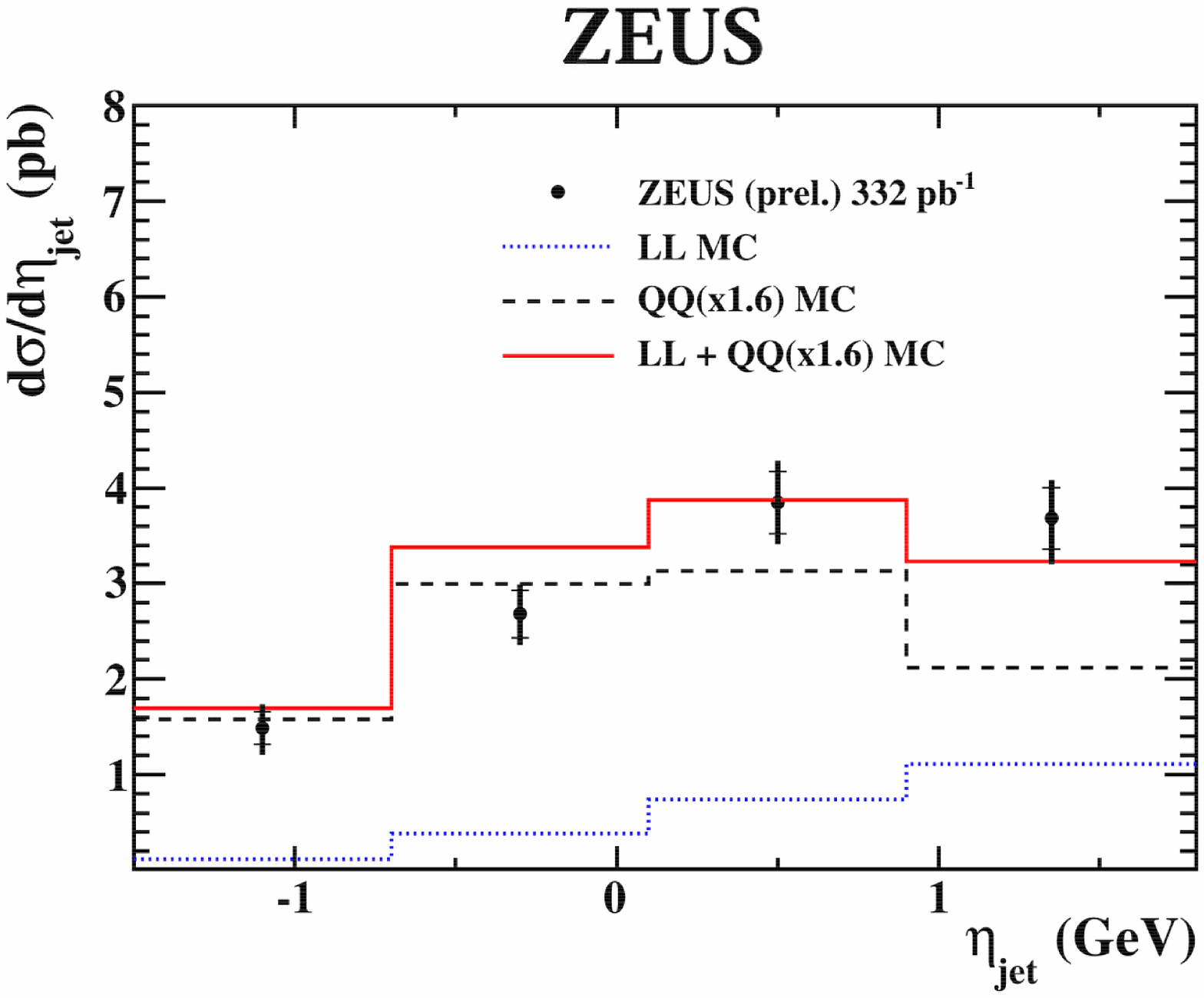}\\[-113mm]
\includegraphics[width=.45\textwidth]{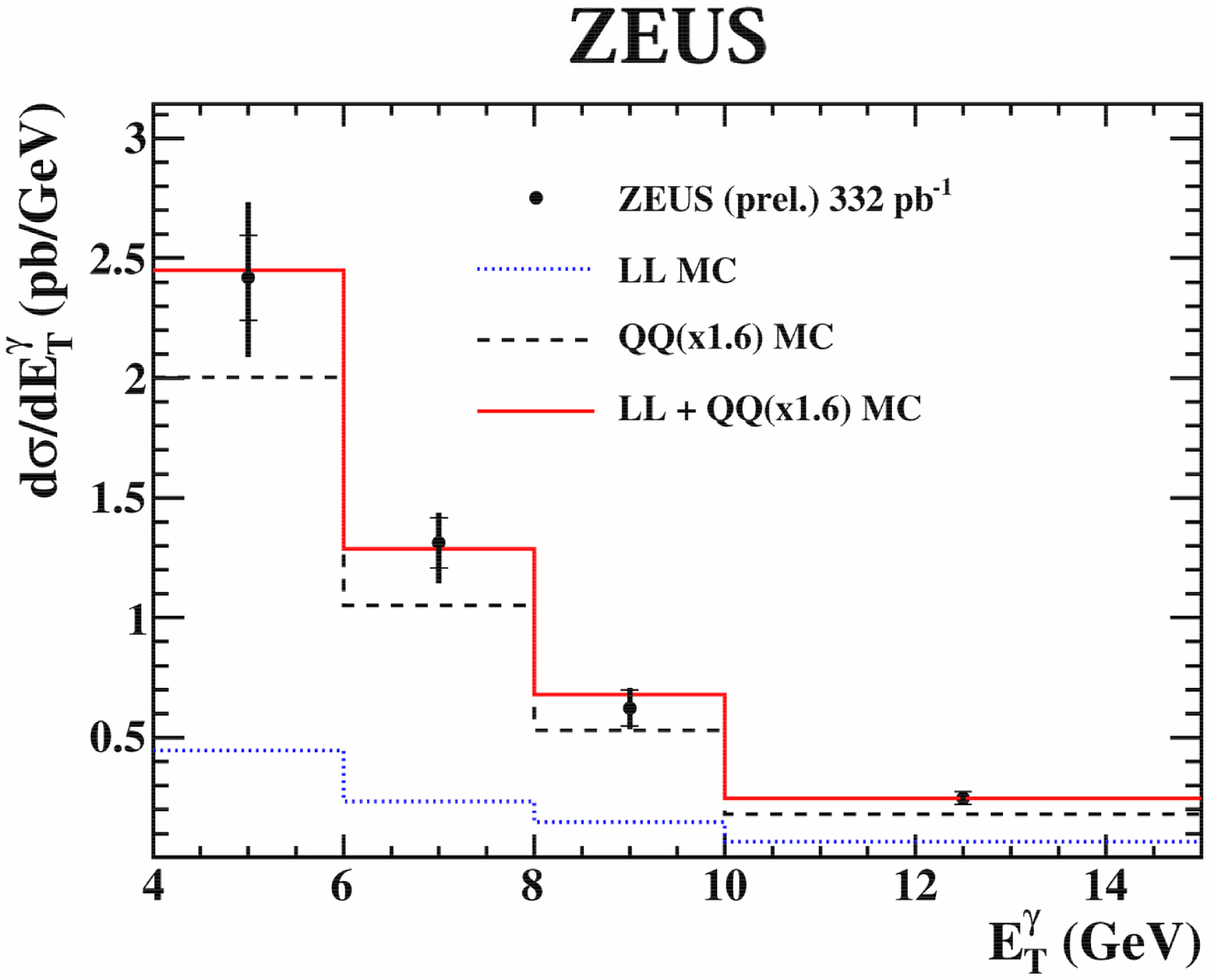}
\includegraphics[width=.45\textwidth]{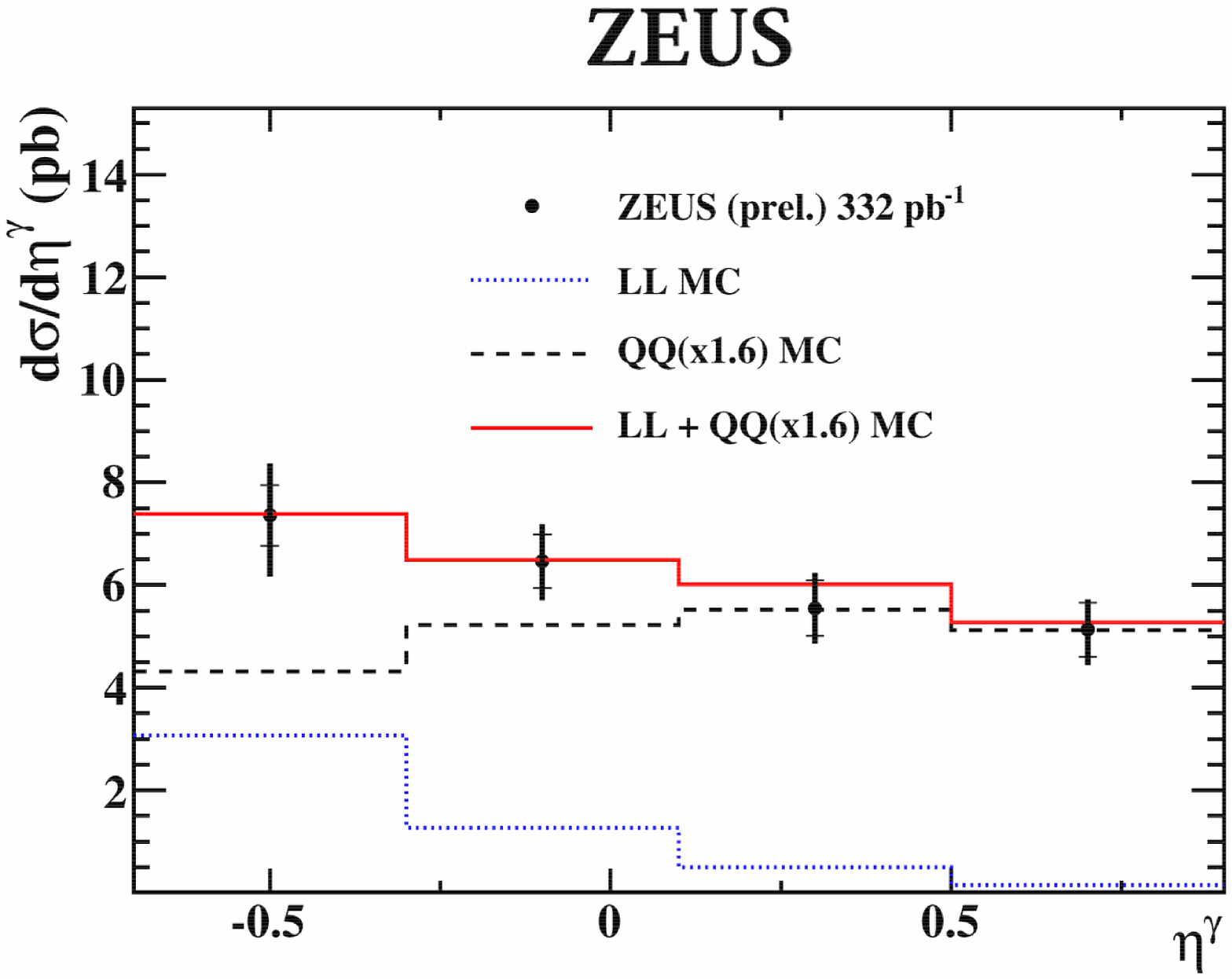}\\[-113mm]
\includegraphics[width=.45\textwidth]{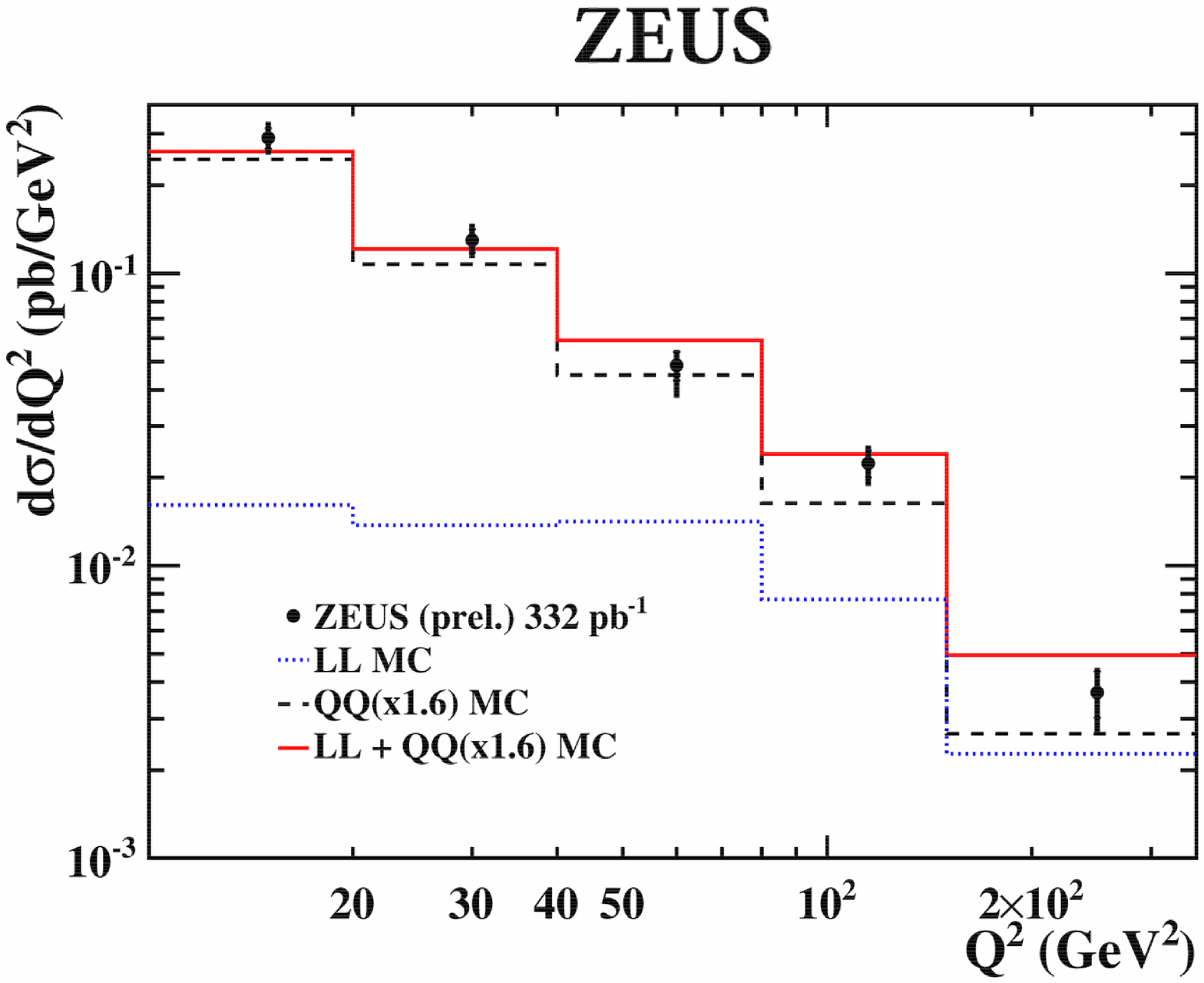}
\includegraphics[width=.45\textwidth]{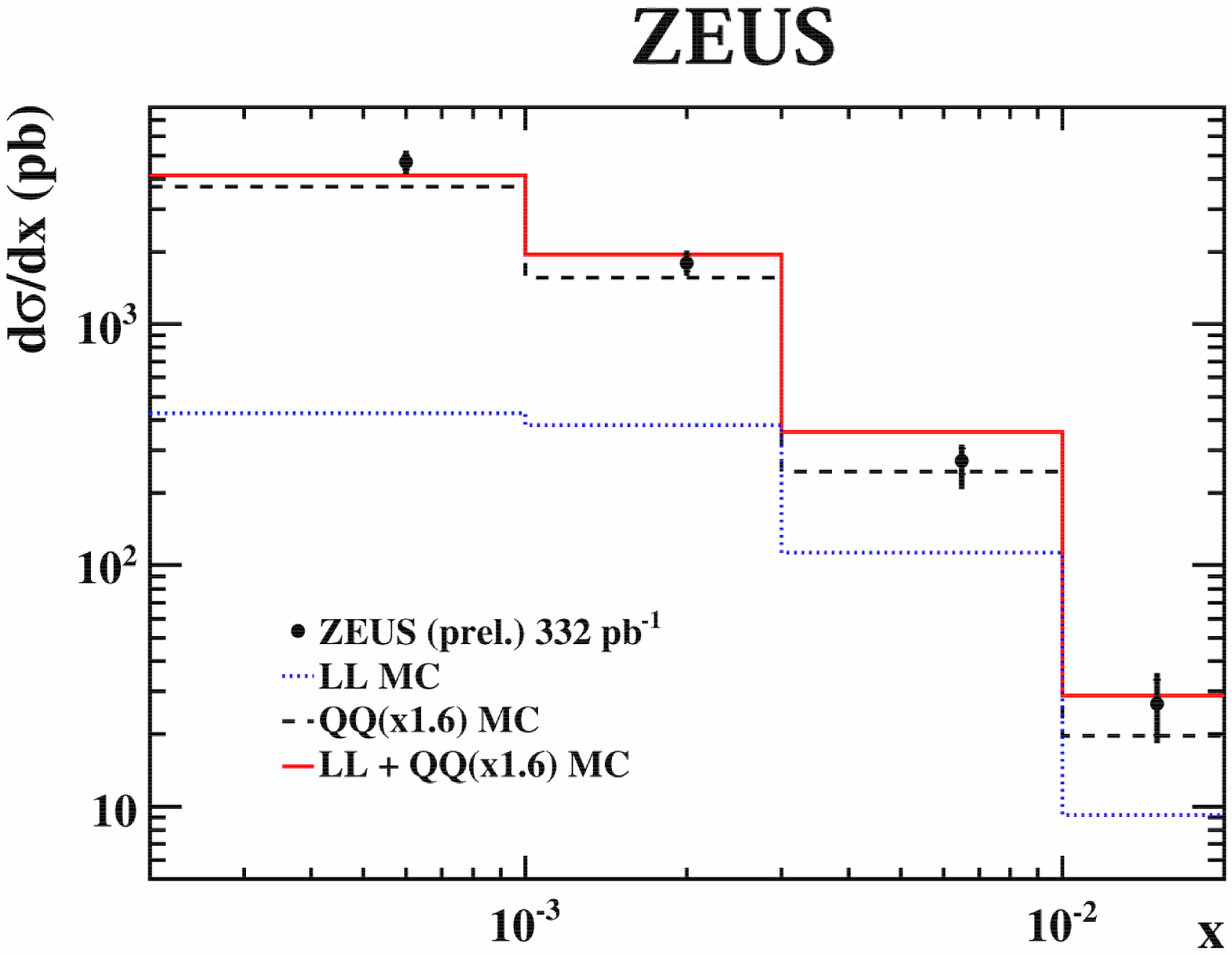}\\[110mm]
\caption{\small Cross sections for kinematic quantities described in
the text, compared to the fitted phenomenological model.}
\end{center}
\end{figure}

\end{document}